\documentclass[twocolumn,prb,showpacs,amsfonts,superscriptaddress,floatfix]{revtex4}
\usepackage{graphicx}
\begin{document}
\title{Luttinger Liquid State with Effective Attractive Hard-Core Interaction}
\author{Igor N. Karnaukhov}
\affiliation{Max-Planck-Institut f\"{u}r Physik komplexer Systeme,
 N\"{o}thnitzer Stra{\ss}e~38, 01187~Dresden, Germany}
\affiliation{Institute of Metal Physics, Vernadsky Street 36, 03142 Kiev, Ukraine}
%\date{today}
\begin{abstract}
An exact solvable 'zig-zag' ladder model of degenerated spinless
fermions is proposed and solved exactly by the means of the Bethe
ansatz. An effective attractive hard-core interaction and direct
Coulomb repulsion of fermions on the nearest-neighbor sites of
different chains induce a new phase state of the ladder. We give a
detailed analysis of the exact phase diagram at zero temperature,
that is characterized by two phases at the filling exceeding 2/3:
itinerant and 'frozen' fermions, for which two different species
reside in spatially separate regions. The critical exponents
describing asymptotic behavior of the correlation functions are
calculated using the Bethe ansatz and conformal field theory. It
is shown also, that the density of the magnetization of the
corresponding zig-zag spin-1/2 ladder has a jump equal to 1/3 at
the magnetic field equal to zero.
\end{abstract}
\maketitle

The metallic single-walled carbon nanotubes (SWNT's) represent
one-dimensional (1D) systems (in the sense of their electronic
properties), in which one may expect to observe physical phenomena
characteristic for strong electron correlations. Informer studies
attention has been mostly concentrated on the realization of the
Tomonaga-Luttinger liquid state in these systems. This conclusion
follows from the analysis of their transport properties
\cite{ex1,ex2} and from the direct observation of electronic
states near the Fermi energy in the high-resolution photoemission
experiments \cite{ex3}. The large value of the critical exponent
$\Theta$ ($\Theta =0.3-0.5$ \cite{ex1,ex2,ex3}), extracted from
the experimental results, which describes the asymptotic behavior
of the spectral function near the Fermi energy, indicates that
large density-density correlations dominate in the system. In
other words, SWNT's are 1D electronic systems with strong
repulsive interaction between fermions. The strong correlations
work wonders in such electron liquid, therefore a traditional
Luttinger liquid approach \cite{llt} that is relevant in the case
of a weak electron-electron interaction can not explain the
mechanism of interactions inherent in carbon nanotubes \cite{k}.
It is necessary to use exactly solvable models or numerical
calculations of many-body fermion systems with strong interactions
for the complete understanding of such intriguing behavior of
SWNT's. We shall exploit this conception and develop an approach
based on the exact solvable zig-zag ladder model of spinless
fermions. An effective attractive hard-core and direct repulsive
interactions are intrinsic in the model considered. As it will be
shown below, the interplay of these interactions gives a rise to
the phase diagram, which includes interesting strongly correlated
states of fermions at the high filling. The values of $\Theta$
calculated in the framework of the model proposed are in
qualitative agreement with Ref. \cite{ex1,ex2,ex3}.

The metallic state of SWNT's is realized in the two possible high
symmetry structures for nanotubes, known as zig-zag and 'armchair'
($m=0$ for all zig-zag tubes  while $n=m$ for all armchair tubes
\cite{d}). The coupled chains in the form of the ladders are hard
lattice models of strongly correlated systems exactly solvable in
uncommon cases. Indeed, such systems are frequently frustrated
\cite{kaw}. The models of the coupled spin chains \cite{f} and
spin ladders have been studied intensively \cite {n,kk,ex}.
Nevertheless, there are still many open questions on the gaps in
the spectrum of collective excitations and the behavior of these
systems away from half-filling. In this Letter we consider the
model for zig-zag nanotubes in the form of the zig-zag ladder of
degenerated spinless fermions taking into account the hopping of
fermions along the chains and the interaction between fermions on
the nearest-neighbor lattice sites of different chains. The model
Hamiltonian is given by
\begin{eqnarray}
{\cal H}= - \sum_{j}(c_{j}^{\dagger
}c_{j+1}+c_{j+1}^{\dagger}c_{j})(1-n_{j+\frac{1}{2}})-JN+\nonumber \\
\epsilon_{1}{\rm M}_{1}+\epsilon_{2}{\rm M}_{2}+
\frac{1}{2}J\sum_{j}(n_{j}+n_{j+1})n_{j+\frac{1}{2}},
\end{eqnarray}
where  $c_{j}$ and $c_{j}^{\dagger }$ are operators of spinless
fermions at site {\it j} shifted on a half-lattice constant for
different chains, accordingly  the first and the second chains,
the hopping integral is equal to unity, $\epsilon_1$ and
$\epsilon_2$ define a relative shift of the fermion subbands of
the chains,  $J$ is the repulsion interaction of fermions on the
nearest-neighbor sites of different chains, the particle number
operator for fermions is defined by $n_{j}=c_{j }^ {\dagger
}c_{j}$. The summation extends over all sites of the chains of the
length L, we assume periodic closure. The hopping of fermions
through the sites occupied by fermions of another chain is
forbidden in the hopping term of the Hamiltonian.  The Hamiltonian
conserves the total number of fermions of each chain  ${\rm M}_1$
and ${\rm  M}_2$, ${\rm N}= {\rm M}_1+{\rm M}_2$  is the total
number of particles. At $\epsilon_{1}=\epsilon_{2}$ the chains are
identical and the Hamiltonian (1) is reduced to  the zig-zag
spin-1/2 ladder in the magnetic  field $H=\epsilon_{1}$ ${\cal H}=
- \sum_{j} (S_{j}^{x}S_{j+1}^x
+S_{j}^{y}S_{j+1}^y)(S_{j+\frac{1}{2}}^z +1/2)-H \sum_{j}S_j^z+
\frac{1}{2}J\sum_{j}(S_{j}^z +S_{j+1}^z)S_{j+\frac{1}{2}}^z.$ By
way of illustration of the structure of the zig-zag ladder and
interaction the chains are shown in Fig.~1.

\begin{figure}[tbph]
\centering{\leavevmode}
\includegraphics[width=2.2in]{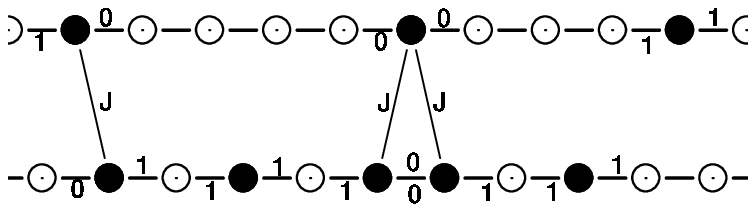}
\caption{Graphical representation of the zig-zag ladder; 0 and 1
denote the value of the  hopping integral along the chains between
the nearest-neighbor sites at given configuration of fermions.}
\label{Fig.1}
\end{figure}

We now turn to the diagonalization of the model Hamiltonian by the
coordinate Bethe ansatz formalism. The Schr\"{o}dinger equation is
solved with the ansatz
\begin{eqnarray}
\psi (x_1,x_2,...,x_N) =
\sum_{P}(-1)^P A(P)\exp\left( i\sum_{j=1}^{N}k_{Pj}x_j\right),
%\nonumber\\
\end{eqnarray}
where the $P$ summation extends over all the permutations of the
momenta $\{k_j\}$ of the particles.

According to the form of the interaction in (1) the fermions of
one chain are scattered on  the fermions of another only, the
two-particle scattering matrix of the spinless fermions with
momenta $k_i$ and $k_j$ is given by $S_{ij} = \exp[\frac{i}{2}(k_i
-k_j)]\frac{1+ \exp [i(k_i+k_j)] +J\exp(ik_j)} {1+\exp[i(k_i+k_j)]
+J\exp(ik_i)}$. An additional constrain on the form of the
interaction in (1) follows from the calculations of the
configurations with three particles in different chains on the
neighboring lattice sites (it is sketched in Fig.~1). Direct
calculations show that if the model Hamiltonian (1) does not
contain a direct interaction of fermions on the nearest-neighbor
sites of the same chain (when the site between these particles is
occupied), the relation between amplitudes of the three-particle
wave function holds for arbitrary configuration of the particles
$A(k_{P'1},k_{P'2},k_{P'3})=S(k_{Pj},k_{Pj+1})A(k_{P1},k_{P2},k_{P3})$
(here $P$ is an arbitrary permutation and $P'=P(j,j+1)$).

The energy eigenstates are characterized  by sets of the charge
rapidities $\{\lambda_j\} (j=1,...,{\rm N})$  for the particles,
that satisfy the following Bethe equations written below in the
case of a weak interaction $J/2=\cos \eta <1$ for the charge
rapidities of the 1-st chain  ($\{\lambda_j \}$,  $j=1,...,M_1$)
\begin{eqnarray}
&&\left [ \frac{\sinh \frac{1}{2}(\lambda_j +i\eta)} {\sinh
\frac{1}{2} (\lambda_j -i\eta)}  \right ]^{{\rm L}+\frac{1}{2}{\rm M}_2} =
\exp\left(\frac{1}{2}{\cal P}_2\right )\nonumber \\
&&
\prod_{i =1}^{{\rm M}_2}\frac{\sinh
  \frac{1}{2}(\lambda_{j}-\lambda_i+2i\eta )}
{\sinh \frac{1}{2}(\lambda _{j}-\lambda_i - 2i\eta )},
\end{eqnarray}
where $\exp(ik_j)=\frac{\sinh\frac{1}{2}(\lambda_j+i\eta)}
{\sinh\frac{1}{2}(\lambda_j-i\eta)}$, ${\cal P}_2
=\sum_{i=1}^{{\rm M}_2}k_i$ is the momentum of the 2-nd chain. The
eigenenergy of the ladder is defined by $E= -2 \sum_{j=1}^N \cos
k_j + \epsilon_1 {\rm M}_1 + \epsilon_2 {\rm M}_2$. The charge
rapidities of the 2-nd chain are solutions of the similar Bethe
equations.

The fillings of the fermionic subbands are defined by the chemical
potential of the system, which is the same for different subbands
at their partial filling, thus  ${\rm M}_1={\rm M}_2$ at
$\epsilon_1 = \epsilon_2$ for identical chains. In the
thermodynamic limit the rapidities $\lambda_j $  are closed spaced
and may be regarded as continuous variable. The distribution
functions of the charge rapidities $\rho_{1,2} (\lambda)$ for each
chain  are defined through the integral equations of the Fredholm
type
\begin{eqnarray}
\rho_{1,2}(\lambda) +
\int_{-\Lambda_{2,1}}^{\Lambda_{2,1}}
d\lambda' R_2 (\lambda -\lambda')\rho_{2,1} (\lambda')=
\nonumber \\
\left( 1+m_{2,1}/2 \right)R_1(\lambda),
\end{eqnarray}
with the kernel
$R_n(\lambda)=\frac{1}{2\pi}\frac{\sin(n\eta)}{\cosh \lambda -\cos(n\eta)}$,
$m_{1,2}=M_{1,2}/L=\int_{-\Lambda_{1,2}}^{\Lambda_{1,2}}
d\lambda \rho_{1,2} (\lambda)$  are the densities of fermions in
the chains, $\varepsilon_{1,2}={\rm
  m}_{1,2}\epsilon_{1,2} - 4 \pi \sin \eta \int_{-\Lambda_{1,2}}^{\Lambda_{1,2}}
d \lambda R_{1}(\lambda)\rho_{1,2}(\lambda)$ are the densities of
the ground-state energy of the chains. The similar Bethe equations
take place in the case of a strong repulsive interaction at $J/2
=\cosh \mu \geq 1$. The corresponding integral equations are
defined by the kernel
$R_n(\lambda)=\frac{1}{2\pi}\frac{\sinh(n\mu)} {\cosh(n\mu)
-\cos\lambda}$.

The low-density region of the phase diagram, when the distance
between particles is large compared to the hard-core radius, describes itinerant
fermions in the Coulomb potential $J$, since an effective force of the the hard-core
interaction is insignificant. Complicated physical picture arises in
region of high densities of fermions, since radically different situation
is realized in the model at high filling
of the chains when the average distance between particles is compared
with the hard-core radius and the hard-core interaction dominates. The
integral equations (4) are defined for the  density of particles less
than a  'half-filling' equaled to 2/3 for identical chains. For
investigation of fermionic states of the ladder at larger filling let us
use the combined electron-hole symmetry $c_j \to (-1)^j c_j^\dagger$
for the Hamiltonian (1) and obtain ${\cal H} \to {\cal  H}'$
\begin{eqnarray}
{\cal H}'=  \sum_{j}[-c_{j}^{\dagger}c_{j+1}-c_{j+1}^{\dagger}c_{j} +\frac{1}{2}J(n_{j}+
n_{j+1})]n_{j+\frac{1}{2}}\nonumber \\
-J{\rm N}+\epsilon_{1}(L-{\rm M}_{1}) +\epsilon_{2}(L-{\rm M}_{2}).
\end{eqnarray}
The Hamiltonian (5) describes the hole states in the ladder,
whereas the fermion states of the first and hole states of the second
chains are defined by the following Hamiltonian
\begin{eqnarray}
{\cal H}'=-\sum_{j=1}^L(c_{j}^{\dagger}c_{j+1}+c_{j+1}^{\dagger}c_{j})n_{j+\frac{1}{2}}
+(\epsilon_1+J) M_1 -\nonumber \\
\sum_{j=1/2}^L(c_{j}^{\dagger}c_{j+1}+c_{j+1}^{\dagger}c_{j})(1-n_{j+\frac{1}{2}})
+\nonumber \\
(\epsilon_{2}-J)(L-{\rm M}_{2})-\frac{1}{2}J\sum_{j}(n_{j}+n_{j+1})n_{j+\frac{1}{2}}.
\end{eqnarray}
\begin{figure}[tbph]
\centering{\leavevmode}
\includegraphics[width=2.2in]{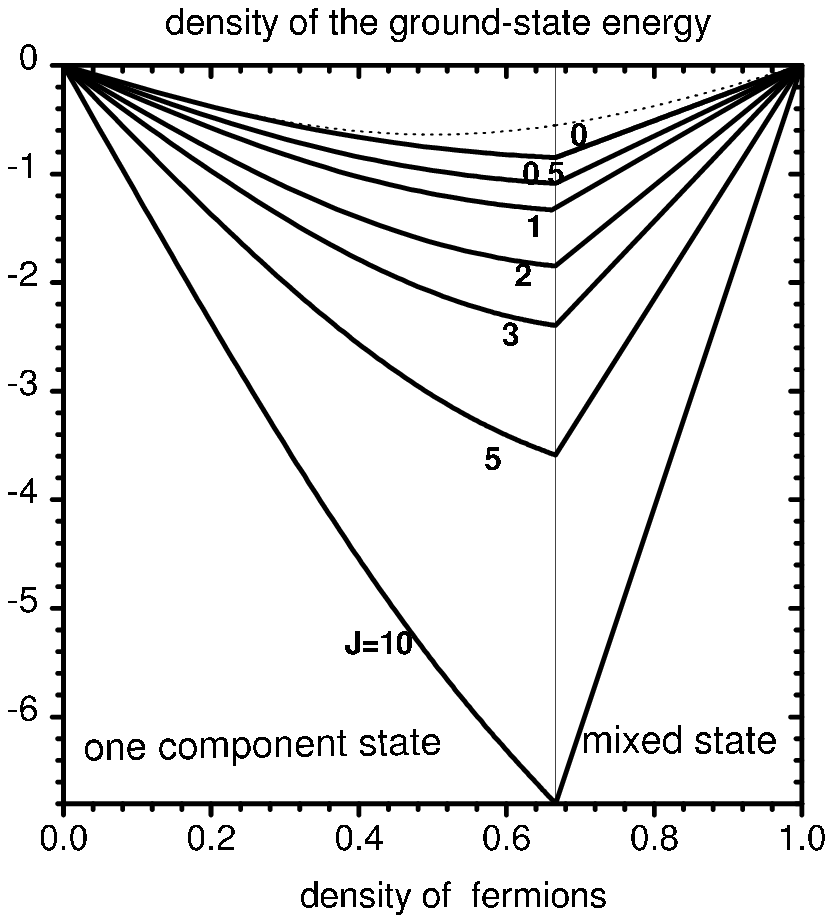}
\caption{Ground-state energy as a function of the density of
spinless fermions. Individual curves are labelled by value of
J=0;0.5;1;2;3;5;10, a dotted curve corresponds to free fermion
state.} \label{Fig.2}
\end{figure}
\begin{figure}[thbp]
\centering{\leavevmode}
\includegraphics[width=2.2in]{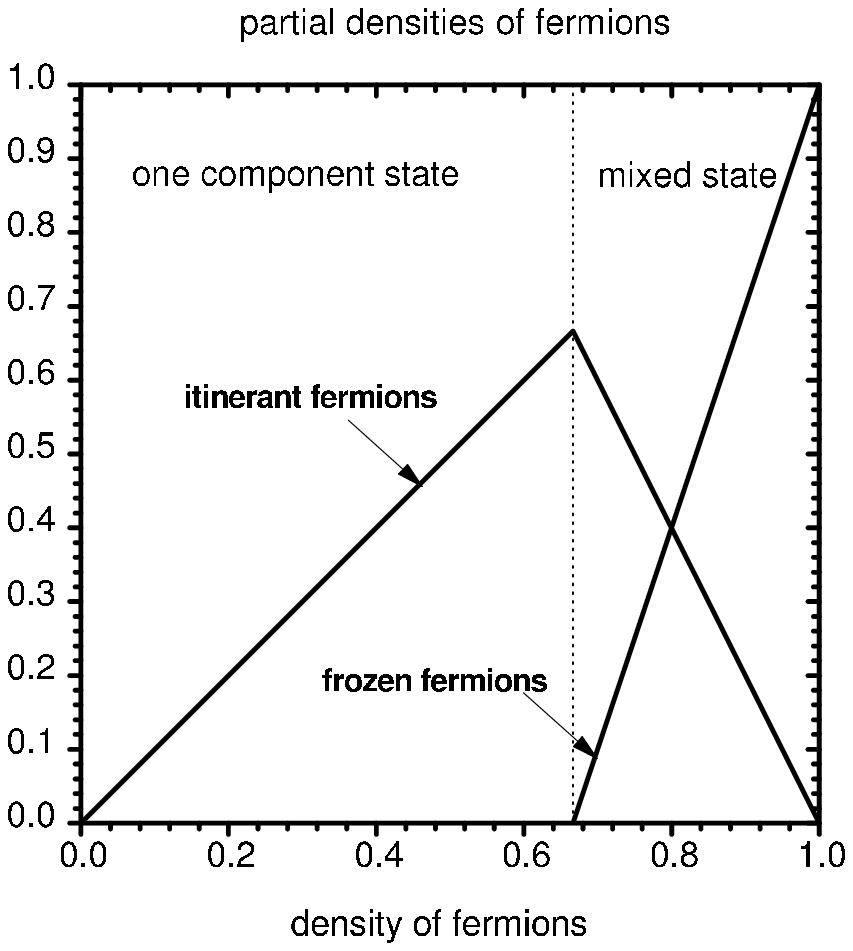}
\caption{Phase diagram in the coordinates of the densities of itinerant
and frozen fermions as a function of the density of fermions.}
\label{Fig.3}
\end{figure}

The phase state of the model is result of the interplay of the repulsive Coulomb
interaction $J$ and effective attractive hard-core interaction with a negative
hard-core radius equaled  to -1/4. We shall show, that phase separation
occurs at high  densities of particles ($n>2/3$).
First of all let us answer on the question: {\it what is it the effective attractive hard-core
interaction?}  A repulsive hard-core interaction forbids two particles being
at distances less or equal to doubled hard-core radius $l$, that corresponds
to equaled to zero two-particle wave function $\psi (x_{i},x_j)$ at
$|x_j -x_i| \leq  l$. The interaction in the Hamiltonian (1) forbids
the tunneling of particles of different chains on distance of a
haft-lattice constant, other words the two-particle wave function is equal
to zero after the result of the hopping (or the current is equal to zero), then
$\psi (x,x) \neq 0$  automatically in the case for a nontrivial
solution  for the wave function. Thus this condition on the two-particle
wave function leads to an effective attraction between two
particles of different chains with the same coordinates in contrast to a traditional repulsive
hard-core interaction for that $\psi (x_{j},x_j)=0$.

Using exact solutions of the Hamiltonians (5),(6) we can calculate
the phase diagram at arbitrary fillings of the chains, below we
consider identical chains. The Hamiltonians (5) is defined the
phase state of fermions with frozen hopping. The Bethe function
(2) is the solution of the Schr\"{o}dinger equations with equaled
to zero energy and the two-particle scattering matrix ${S}_{ij}'
=\exp[\frac{i}{2}(k_j -k_i)]$. Really this phase corresponds to
full-filled states of the chains, other words these solutions
describe the phase separation of the chains at $n>2/3$: the
full-filled states of fermions in the chains with frozen hoppings
and the 'old' phase of itinerant fermions defined by the Bethe
equations (3) on the restricted chains. The density of the
ground-state energy is plotted in Fig.~2 at
$\epsilon_1=\epsilon_2=0$. At $n>2/3$ the density of 'frozen'
fermions is equal to $\delta n =3n-2$, the corresponding density
of the itinerant fermions $\Delta n = n-\delta n =2(1-n)$ (see
Fig.~3). In a separate point $J\to \infty$ the scattering matrix
is defined by conditions $\psi (x_{j \pm \frac{1}{2}},x_j)=0$,
that corresponds to a constrain on the tree-particle wave function
$\psi (x_{j+1},x_{j+\frac{1}{2}},x_j)=0$ also, therefore the
maximal filling of chains is equal to 1/3. In the case of a strong
repulsive interaction $J>2$ both phases are dielectric, the phase
of itinerant fermions is metallic for a weak interaction $J \leq
2$.

The Luttinger liquid parameter $K_\rho^{1,2}$ is defined by the
dressed charge $\chi_{1,2} (\Lambda)$ via $K_\rho^{1,2}=\chi_{1,2}^2 (\Lambda)$.
The integral equations for the dressed charge functions $\chi_{1,2}
(\lambda)$ is distinguished from equations (4) by the driving term
\begin{eqnarray}
\chi_{1,2}(\lambda) +\int_{-\Lambda_{2,1
}}^{\Lambda_{2,1}}
d\lambda' R_2 (\lambda -\lambda')\chi_{2,1} (\lambda')=
%\nonumber \\
\left( 1+m_{2,1}/2\right).
\end{eqnarray}
\begin{figure}[thbp]
\centering{\leavevmode}
\includegraphics[width=2.2in]{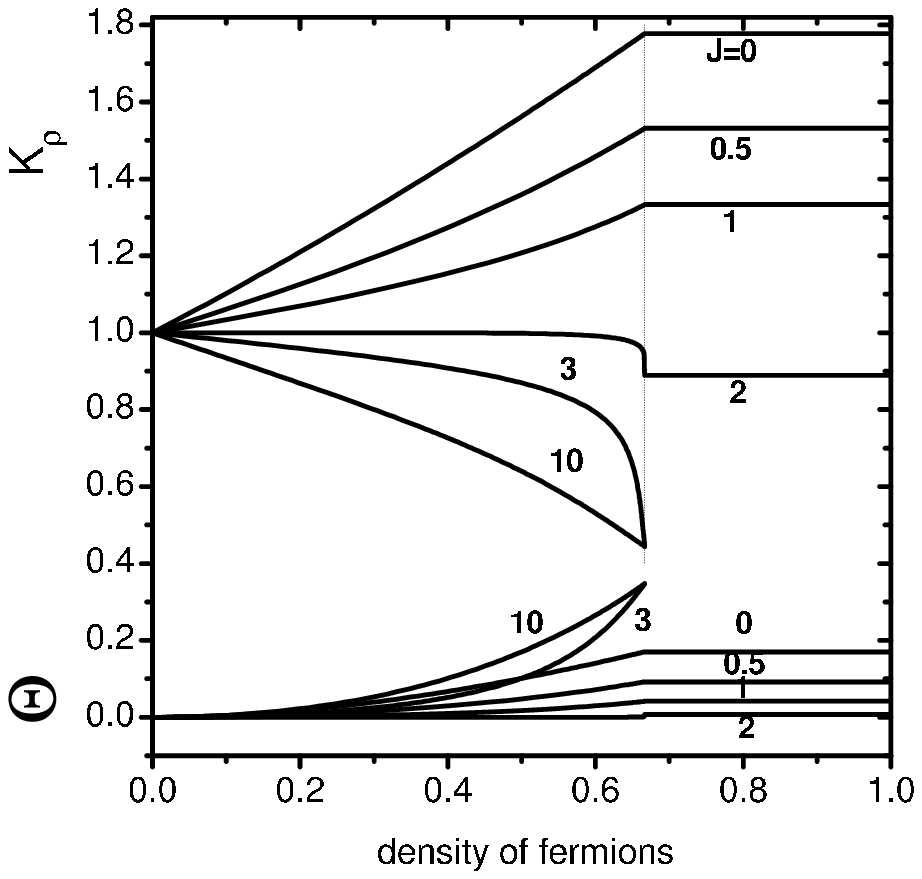}
\caption{The critical exponents $K_\rho$ and $\Theta$ at J=0;0.5,1;2;3;10.}
\label{Fig.4}
\end{figure}
\begin{figure}[tbph]
\centering{\leavevmode}
\includegraphics[width=2.2in]{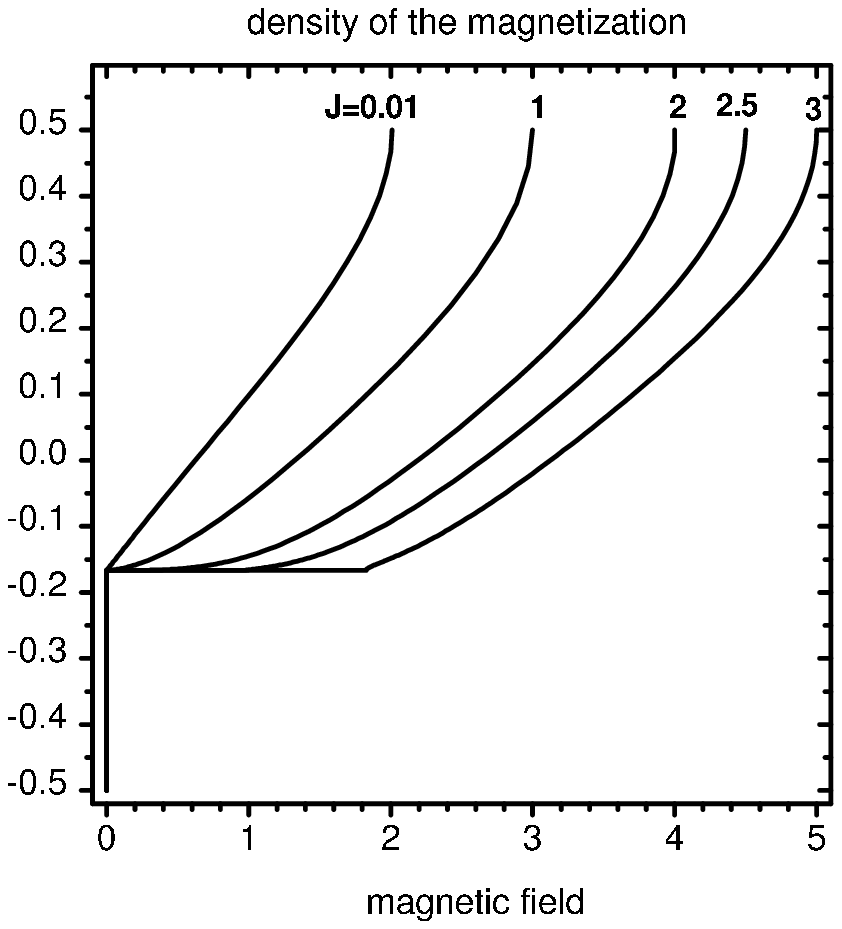}
\caption{Density of the magnetization of the spin ladder
model as a function of the magnetic field at J=0.01; 1; 2; 2.5; 3. }
\label{Fig.5}
\end{figure}

According to (7) the repulsive interaction $J$ decreases the value
of $K_\rho$. In contrast to traditional hard-core interaction,
that decreases the value of  $K_\rho$ extremely \cite{k}, an
effective attractive hard-core interaction  increases $K_\rho$.
The low density limit $m_{1,2} \to 0$ corresponds to free spinless
fermions with $K_\rho \to 1$. In the case of the identical chains
the system of integral equations (7) is reduced  to one and we
obtain the solution for $K_\rho =\frac{8}{9}\frac{\pi}{\pi -\eta}
$ at half filling $m_{1,2}=2/3$ using the Wiener-Hopf method. The
maximal value of $K_\rho =16/9$ is realized in the $J\to 0$ limit,
that corresponds to $\Theta=0.17$. The critical exponent $\Theta$
is related to the Luttinger parameter $K_\rho$ through $\Theta
=(K_{\rho}-2+1/K_{\rho})/2$. Whereas, the minimal value $K_\rho
=4/9$ is obtained in the $J \to \infty$ limit, we have twice
larger value of the critical exponent $\Theta =0.35$ in this case.
Numerically solving integral equations (4),(7) we calculate the
critical exponents $K_{\rho}$ and $\Theta$ for each chain as
function of the density of spinless fermions for different values
of $J$.  Numerical calculations of the exponents  are shown in
Fig.~4 for illustration of the behavior of the critical exponents
for arbitrary densities of particles and interaction. In an
effective attractive regime at a weak interaction ($J<2$) the
value of $K_\rho$ decreases from 16/9 to 1 with increasing J from
0 to $J =J_0$ (the value of $K_\rho$ depends of the density
fermions therefore $J_0$ depends of $n$ also, according to
simulations $1.7< J_0 <1.9$), whereas the value of $\Theta$
decreases from 0.17 to zero. The phase of itinerant fermions is
metallic at $n>2/3$, therefore the critical exponents independent
of the density fermions at $n>2/3$. For $J>J_0$ $K_\rho$ decreases
with $n$ and reaches a minimal value 4/9 at half-filling, the
value of $K_\rho$ decreases with $J$ also. The value of $\Theta$
increases and reaches a maximum value 0.37 at half-filling, both
phases are dielectric for $J>2$ at $n \geq 2/3$. At half-filling
the value $J=2$ separates the phase of itinerant fermions on
gapless at $J\leq 2$ and gaped at $J>2$. The behavior of the
density of the magnetization as a function of the magnetic field,
calculated for different values of the interaction J, is shown in
Fig.~5. The magnetization of the system has a jump at H=0 equal to
$1/3$. This jump corresponds to the phase of frozen fermions with
the chemical potential equal to zero.

In summary, we presented the exact phase diagram of the zig-zag
ladder model for arbitrary Coulomb repulsion interaction and
density of fermions obtained on the basis of the exact solution of
the model. It is shown that intrinsic effective attractive
hard-core interaction and Coulomb repulsion form the Luttinger
liquid state in the ladder. Due to special role of the effective
attractive hard-core interaction,  two phases of the itinerant and
frozen fermions are separated at the high density, when filling
exceeds 2/3. As a result, the system exhibits two types of
behavior in high density region: either band insulator in the case
of the strong interaction at $J>2$ or the Luttinger liquid state
for one phase and band insulator for another at $J\leq 2$. The
values of bulk exponents calculated in the framework of the model
adopted are in a reasonable agreement with the measurements in
SWNT's.

The author wishes to thank the support of the Visitor Program
of the Max-Planck-Institut f\"{u}r Physik Komplexer  Systeme, Dresden,
Germany. Part of this work was performed at the International Centre
for Theoretical Physics, Trieste, Italy. Research supported by the Nano
Structure Systems, Nano materials and Nano Technology Program, project STSU-3520.

%\begin{references}

%\end{references}

\begin{thebibliography}{31}
\bibitem{ex1}H. Ishii {\it et al.,} Nature (London) {\bf 426}, 540 (2003).
\bibitem{ex2}   A. Bachtold, M. de Jonge, K. Grove-Rasmussen, P. L. McEuen,
M. Buitelaar and C. Sch�enberger, Phys. Rev. Lett. {\bf 87}, 166801 (2001).
\bibitem{ex3}M. Bockrath {\it et al.,} Nature (London) {\bf 397}, 598
(1999).
\bibitem{llt} R. Egger and A.O. Gogolin, Eur. Phys. J. B {\bf 3}, 281
(1997); R. Egger, Phys. Rev. Lett. {\bf 83}, 5547 (1999); C. Kane,
L. Balents, and M.P.A. Fisher, Phys. Rev. Lett. {\bf 79}, 5086 (1997).
\bibitem{k}I.N. Karnaukhov, V.I. Kolomytsev, and C.G.H. Diks,
  Phys. Rev. B {\bf 75}, 125407 (2007); I.N. Karnaukhov and
  A.A. Ovchinnikov, Europhys. Lett. {\bf 57}, 540 (2002).
\bibitem{d}R. Saito, G. Dresselhaus, and M.S. Dresselhaus, {\it Physical
  Properties of Carbon Nanotubes} (Imperial College Press, London,
  1998).
\bibitem{kaw} Z. Wiehong, V. Kotov, and J. Oitmaa, Phys.Rev. B {\bf 57},
  11439 (1998); X. Wang, Mod. Phys. Lett. B {\bf 14}, 327 (2000);
  T. Hakobyan, Phys.Rev. B. {\bf 75}, 214421 (2007).
\bibitem{f} H. Frahm, M. Pfannm\"{u}ller, and T.M. Tsvelik, Phys. Rev. Lett.
  {\bf 81}, 2116 (1998).
\bibitem{n} A.A. Nersesyan and A.M. Tsvelik, Phys. Rev. B  {\bf 68}, 235419 (2003).
\bibitem{kk} D.N. Aristov, M.N. Kiselev, and K. Kikoin, Phys. Rev. B {\bf 75}, 224405 (2007).
\bibitem{ex}Y. Wang, Phys. Rev. {\bf 60}, 9236 (1999);
M.T. Batchelor and M. Maslen, J.Phys. A {\bf 32}, L377 (1999);
H. Frahm and C. R\"{o}denbeck, J. Phys. A., {\bf 30}, 4467 (1999);
H. Frahm and A. Kundu, J.Phys. Condens. Matter {\bf 11}, L557 (1999);
P. Corboz, A.M. L\"{a}uchli, K. Totsuka, and H. Tsunetsugu /
cond-mat.0707.1195 (9 July 2007).
\end{thebibliography}
\end{document}